# Enhanced strength and temperature dependence of mechanical properties of Li at small length scales and its implications for Li metal anodes

Chen Xu[a,*], Zeeshan Ahmad[b], Asghar Aryanfar[a], Venkatasubramanian Viswanathan[b], and Julia R. Greer[a]

[a]Division of Engineering and Applied Science, California Institute of Technology, Pasadena, CA 91125

[b]Department of Mechanical Engineering, Carnegie Mellon University, Pittsburgh, PA 15213

Address correspondence to: chenxu@caltech.edu; venkvis@cmu.edu

**Most next-generation Li-ion battery chemistries require a functioning lithium metal (Li) anode. However, its application in secondary batteries has been inhibited because of uncontrollable dendrite growth during cycling. Mechanical suppression of dendrite growth through solid polymer electrolytes (SPE) or through robust separators has shown the most potential for alleviating this problem. Studies of the mechanical behavior of Li at any length scale and temperature are virtually non-existent because of its extreme reactivity, which renders sample preparation, transfer, microstructure characterization and mechanical testing prohibitively challenging. We conduct nano-mechanical experiments in an *in-situ* Scanning Electron Microscope and show that micron-sized Li attains extremely high strengths of 105 MPa at room temperature and of 35MPa at 90ºC. We demonstrate that single crystalline Li exhibits a power-law size-effect at the micron- and sub-micron length scales, with the strengthening exponent of -0.68 at room temperature and of -1.00 at 90 ºC. We also report the elastic and shear moduli as a function of crystallographic orientation gleaned from experiments and first principles calculations, which show a high level of anisotropy up to the melting point, where the elastic and shear moduli vary by a factor of ~4 between the stiffest and most compliant orientations. The emergence of such high strengths in small-scale Li and sensitivity of this metal's stiffness to crystallographic**

**orientation help explain why the existing methods of dendrite suppression have been mainly unsuccessful and have significant implications for practical design of future-generation batteries.**

Increased adoption of electric vehicles (EV) requires an improvement in the energy density of rechargeable Li-ion batteries. Li metal anode is a common and necessary ingredient in commercialization pathways for next generation Li-ion batteries. In the near term, Li metal coupled with an advanced cathode could lead to a specific energy of 400 Wh/kg at the cell level, which represents 200% improvement over current state of the art[1]. In the longer term, Li metal coupled with a S and $O_2$ cathode could lead to even higher specific energies of > 500 Wh/kg. Despite over 40 years of research, overcoming the uncontrollable dendrite growth during cycling has remained an insurmountable obstacle for Li-based components[2]. Among multiple attempted approaches to eliminate or even reduce the dendrite growth, mechanical suppression has emerged as one of the most promising routes. The pioneering theoretical work by Monroe et al. showed that SPEs whose shear modulus is roughly twice that of Li could suppress dendritic growth through compressive forces[3]. This led to enormous interest in demonstrating cells with polymer electrolytes, inorganic solid state Li-ion conductors and ceramic thin films[2]. Ferresse et al. further demonstrated that the elastic modulus of the separator also affects dendrite growth because it causes the stress in the separator to build up to beyond the yield strength of Li, which causes the anode to plastically deform and flatten out[4]. Applying external pressure higher than the yield strength of bulk polycrystalline Li in a direction perpendicular to the cell stack has also been shown to limit dendrite growth and prolong cycle life[5–7].

These approaches have had limited success, and many unsolved questions regarding mechanical suppression remain. For example, dendrites form and grow through the grain structure of Li garnet solid electrolytes even though their shear modulus is >50GPa, a value predicted to be sufficiently high to suppress dendrite growth[8]. Applying an external pressure above what is believed to be the yield strength of Li also does not fully eradicate dendrites most likely because of the dearth of high-fidelity mechanical properties data for Li. Elastic modulus of polycrystalline Li has been reported to range from 1.9 GPa to 7.9 GPa[9–12], and its yield strength – from 0.41 MPa to 0.89 MPa[11,12]. Such a significant variation in both has been attributed to the differences in sample preparation methods, i.e. melting and annealing conditions, reactivity with atmosphere, and experimental error. A key reason for the lack of solutions to overcome the Li dendrite growth challenge may be that the mechanical properties of Li at small scales are expected to drastically differ from those in its bulk form; most single crystalline metals at the micron- and sub-micron scales have been shown to be up to an order of magnitude stronger compared with their bulk form[13]. During charging, crystalline Li whiskers with diameters on the order of a few hundred nanometers have been observed to nucleate on the anode surfaces[14], growing into dendrites up to several millimeters long. The size-*independent* properties, such as the elastic and shear moduli, are generally functions of crystallographic orientation, and are particularly sensitive to it in Li, whose anisotropy factor is 8.52 at room temperature[15]. The mechanical properties of Li at high temperatures are also largely unknown. Current SPEs require an operating temperature of 333 K to 363 K (60 °C to 90 °C) to achieve the desired ionic conductivity ($10^{-3}$ S/cm) and strong adhesion to the electrodes. Li's low melting temperature of 453 K (180 °C) suggests that even a modest temperature elevation will likely have a dramatic effect on its mechanical properties.

In this work, we aim to fill the gaps in the existing severely incomplete understanding of the mechanical properties of Li. We performed uniaxial compression experiments on single crystalline Li pillars with diameters of 980 nm to 9.45 μm at 298 K and 363 K in an *in-situ* nanomechanical instrument inside of an SEM chamber. We observed a pronounced size effect in yield strength at room temperature, with the power law exponent of strength vs. pillar diameter of -0.68, a value that is higher than those reported for most body-centered cubic metals[13,16]. This size effect becomes more pronounced at 363 K, where the exponent decreases to -1.00. We further observed a dramatic decrease in yield strength across all pillar sizes at 363 K compared with that at room temperature, by a factor of ~3.5. To address the mechanical properties in the elastic regime, density functional theory (DFT) calculations were performed to determine the elastic constants as a function of temperature from 78 K to 440 K and found them to be in good agreement with experimental values, where applicable (78K to 300 K)[17,18]. Size-*independent* properties such as the elastic and shear moduli were calculated as a function of crystal orientation and temperature. We discuss the impact of the discovered size effect and of elastic anisotropy on Li dendrite growth in the framework of nano-crystalline plasticity and provide insights into the shortcomings of the existing dendrite suppression methods, as well as into the potential pathways to utilize these findings in designing safer and more efficient energy storage systems.

**Methods**

Li granules (Sigma Aldrich) were melted on a hotplate in an Argon glovebox at 180 °C, then cooled down to room temperature over 4 hours. The sample was then cut with a surgical blade to reveal a shiny metallic surface, from which we fabricated the nano- and micro-compression samples. This parent Li sample was placed into an airtight transfer module (Vacushut, Agar Scientific) and carried from the glovebox to the SEM without exposure to the atmosphere. Ar was used to vent the SEM. Once inside the SEM chamber, the lid of the Vacushut self-opened as the chamber was pumped down to vacuum. We fabricated cylindrical pillars with diameters ranging from 980 nm to 9.45 μm in a Focused Ion Beam (FIB) (Versa DualBeam, FEI); the aspect ratio, height/diameter, was maintained between 3:1 and 5:1. The tops of the pillars were polished in the FIB to minimize surface roughness. After fabrication, the sample was carried in the Vacushut from the FIB to the *in-situ* nanomechanical instrument, SEMentor (InSEM, Nanomechanics, Inc and FEI) (Fig. S1). SEMentor allows us to simultaneously capture the real-time deformation video and to collect mechanical data during the experiments. Samples were compressed using a custom-made diamond flat punch tip with a diameter of ~12 μm, and all experiments were conducted under a constant nominal strain rate of $5\times10^{-3}$ $s^{-1}$. Engineering stresses and strains were calculated by dividing the applied force and displacement by the initial cross-sectional area and pillar height, respectively. The initial cross-sectional area was calculated based on the SEM images, using the pillar diameter measured at halfway along the pillar height. We accounted for the compliance of the substrate using Sneddon's correction[19]. SEMentor is equipped with a heating module located directly underneath the sample mount, which allows us to heat the sample up to 200 °C. We used a thermocouple located within the sample mount and connected to a PID temperature controller (Lake Shore Cryotronics, Inc) to carefully maintain the sample at the set temperature (Fig. S1). We found that the sample stabilized at 363 K (90 °C) after 3 hours. To avoid significant thermal drift, i.e. recorded displacement caused by thermal expansion of the tip/sample and possible temperature fluctuations during compression, the tip was placed in contact with the lithium substrate for at least 3 hours to equilibrate the temperature before each test. We recorded typical thermal drift after each test to be <5 nm/s, which we used to correct the stress-strain data. After compression, we transferred the Li sample to another SEM (Zeiss 1550VP FESEM) to determine the crystallographic orientations of each pillar using Electron Backscatter Diffraction (EBSD). This was done post-compression because the Vacushut is not compatible with the Zeiss SEM, and therefore the sample must be removed from the Vacushut and temporarily exposed to air before doing EBSD. In the course of this work, we learned that even a 30 second exposure to the atmosphere causes severe oxidation in Li, which manifests itself in a dramatically different mechanical response (Fig. S2). Fig. S3 shows that surface contamination induced by the transfer lowered the indexing rate of EBSD. We employed a new piece of Li every time the above-mentioned sequence of experiments was conducted. The effects of $Ga^+$ implantation during FIB milling on the mechanical properties of Li is investigated and discussed in the SI.

The elastic modulus ($E$) of each pillar was obtained from the unloading section of the stress-strain curve. The top 1/3 of the unloading curve, starting at the point of maximum depth, is linearly fitted and the slope is taken to be $E$. The method of obtaining the crystal orientation using the obtained value for $E$ is given in the SI.

We used Density Functional Theory (DFT) to calculate the orientation- and temperature-dependent elastic constants of Li. We use the quasi-harmonic approximation to account for the free energy contribution from lattice vibrations[20]. Calculations were performed using a real space projector augmented wave (PAW)[21] method as implemented in GPAW[22] within the generalized gradient approximation (GGA)[23]. Details on the calculation method are given in the SI.

**Results**

Fig. 1a shows representative engineering stress vs. engineering strain data for room temperature compression of Li micropillars with different diameters. All curves exhibit an initial elastic loading segment followed by plastic flow, ending in catastrophic failure at a strain of 2 -3 %. The initial loading slope corresponds to the stiffness of each sample; the differences in these slopes are caused by the different crystallographic orientations of the samples, as well as possible effects of slight initial misalignment during the experiments[24]. We observed a pronounced increase in the yield and flow stresses as the pillar diameter was decreased. The yield strength, defined as the stress at which the first significant strain burst occurs, increases from 15 MPa to 105 MPa as the pillar diameter decreases from 9.45 µm to 1.39 µm. Figure 1(b, c) shows snapshots of a 1 µm pillar during compression, where the pillar sheared off via a single slip offset. Figure 1(d,e) shows SEM images of some representative Li pillars deformed at room temperature, with characteristic sharp and localized slip traces.

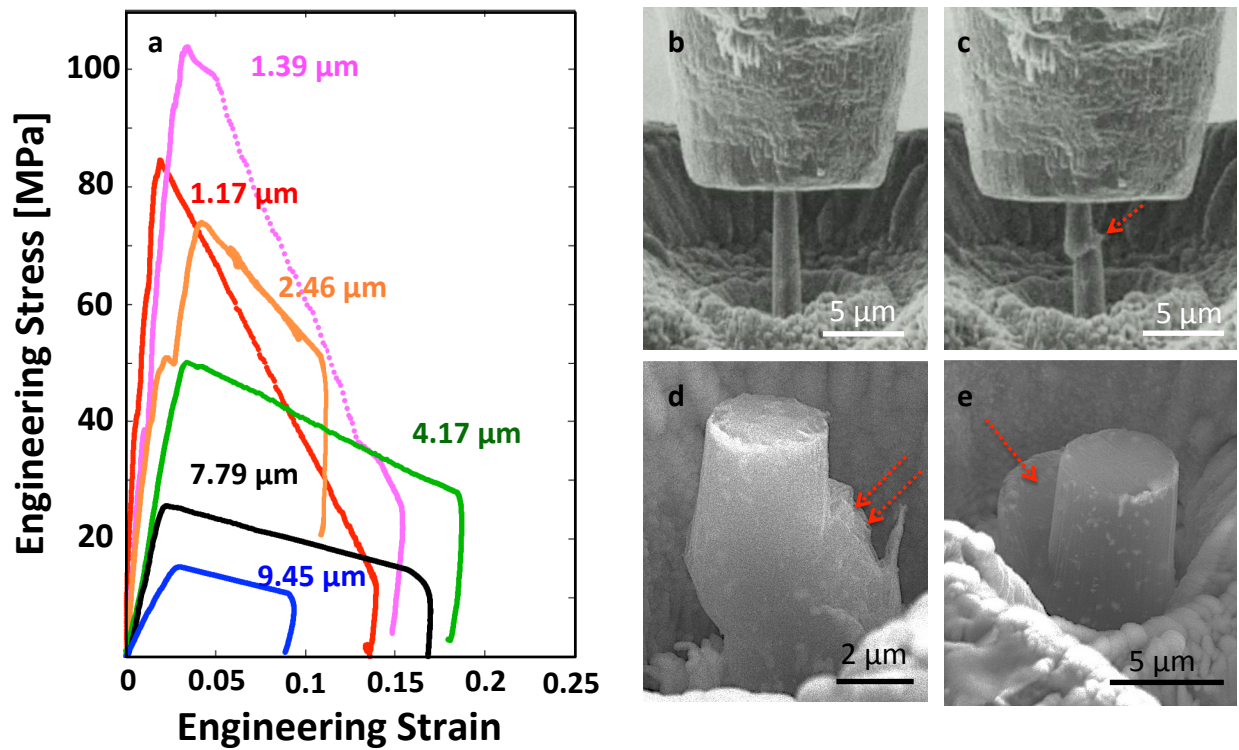

**Figure 1**. **Room-temperature uniaxial compression experiments**. (a) Engineering stress-strain data for Li pillars with different diameters. (b-c) Snapshots of *in-situ* compression of a representative 1 µm-diameter Li pillar. (d,e) Compressed 4 µm–diameter Li pillars. Arrows point to slip offsets, likely along a <111> direction.

We performed another series of micro-compression experiments at 363K. Figure 3(a, b) shows the mechanical response of Li pillars at 363 K compared with room temperature experiments. These plots reveal that the yield strength of ~1μm-diameter samples at 363 K is around 35 MPa, compared to ~95 MPa at 298 K; and for 8.5 μm diameters, the yield strength is 5 MPa at 363 K compared to 16 MPa at 298 K. A modest decrease in yield strength during higher-temperature deformation has been reported for other bcc metals[25–27]; a factor of 3 decrease in yield strength of Li at only 65 K higher temperature discovered in this work has not been observed before. Tariq et al. reported the ultimate tensile strength of bulk polycrystalline Li deformed at a strain rate $2\times10^{-3}$ s$^{-1}$ at room temperature to be 0.89 MPa, and 0.46 MPa at 75 °C[12]. Figure 3(c,d) shows SEM images taken during *in-situ* compression of an 8.5μm-diameter pillar and reveals that, unlike in room temperature experiments, this sample experiences barreling with no visible crystallographic slip offsets.

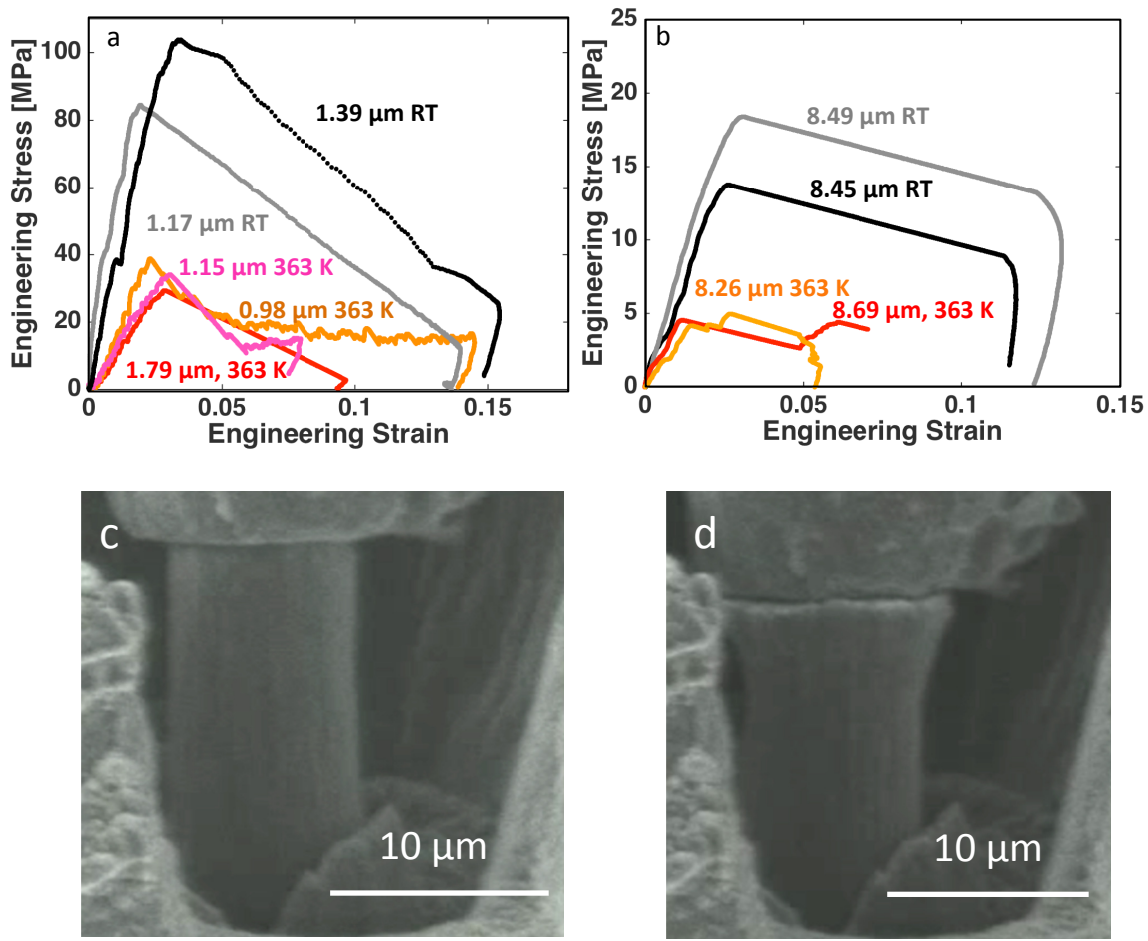

**Figure 2**. Uniaxial compression stress vs. strain for compression of Li pillars at room temperature (RT) and 363 K. (c, d) snapshots of *in-situ* compression of a 9 μm diameter Li pillar at 363 K, which shows barreling and no crystallographic offsets.

Figure 3a shows the plot of Critical Resolved Shear Stress (CRSS) at 2% strain as a function of pillar diameter on a log-log scale. We chose the most common slip system in bcc metals - {100}<111>[28] to calculate CRSS for all experiments by multiplying the axial flow stress at 2% strain by the maximum Schmidt factor allowed for the slip system. The crystallographic orientation of each pillar was estimated either directly from EBSD map or from the unloading data when EBSD mapping was unavailable (details given in the SI). We found the power law slope for size-dependent strengthening of Li at room temperature to be -0.68 and -1.00 at 363 K. Figure 4b shows the CRSS normalized by the shear modulus, $G^3$, as a function of the pillar diameter, $D$, normalized by the Burgers vector, $b^{29}$, for Li and several other BCC metals deformed at room temperature. It shows that at room temperature, Li has the highest relative strength among all metals in this size range. Additionally, the size effect slope for Li deformed at 363 K is higher than virtually all other size effect slopes reported for single crystalline BCC metals[13].

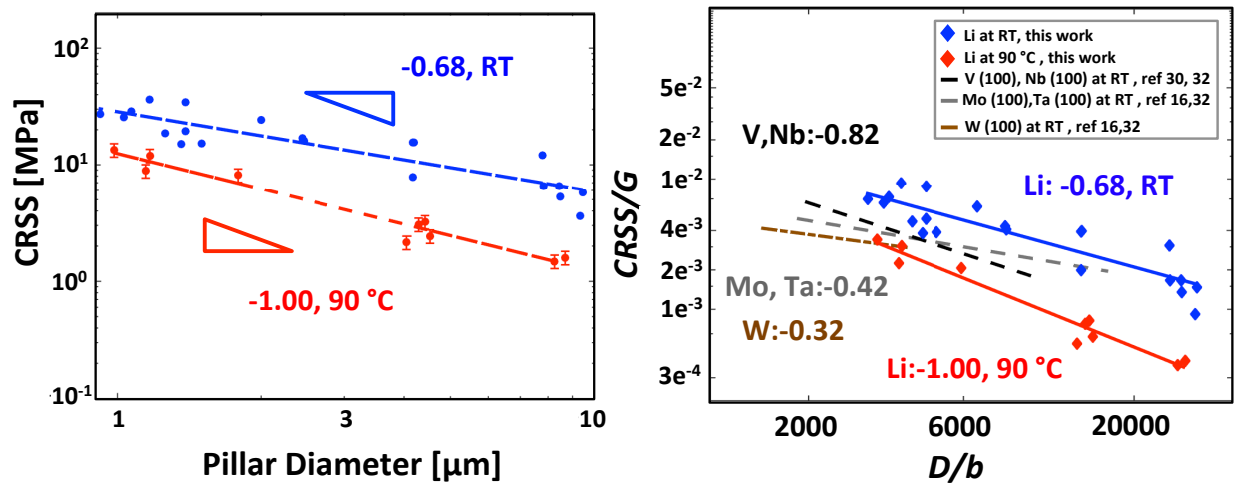

**Figure 3**. **Size effect in Li at RT and 363K compared with other BCC metals.** (a) CRSS vs. pillar diameter for room temperature (RT) and 363 K experiments. Error bars are plotted for those data points where pillars bent after catastrophic failure, causing the unloading data to no longer reflect the stiffness. In those cases, the Schmidt factor may range from 0.500 to 0.272, and the error bars represent one standard deviation. (b) CRSS normalized by the bulk shear modulus vs. pillar diameter normalized by the Burgers vector for different single crystalline bcc metals. The values for the slopes are of the CRSS vs. D, not the normalized values.

In addition to strength, the elastic properties of Li, such as the moduli, which are size-*independent,* are also expected to affect Li dendrite growth[3]. In a single crystal, the elastic properties such as the shear modulus, $G$, elastic modulus, $E$, and Poisson's ratio, $\nu$ vary with crystallographic orientation. Li has a very high anisotropy factor of 8.52, defined as $A = \frac{2C_{44}}{C_{11} - C_{12}}$, where $C_{11}$, $C_{21}$ and $C_{44}$ are elastic constants[17,18]. This implies that the mechanical properties of each Li grain vary significantly with orientation, which leads to a broad range in shear and elastic moduli and different propensity for dendrite formation and growth. Experimental values of elastic constants as a function of temperature exist only in the range of 78 – 300 K[17,18]. We used DFT to calculate the elastic constants of Li between 78 K and 440 K. Figure 4a shows $C_{11}$, $C_{12}$, and $C_{44}$, plotted as a function of temperature along with the experimental data from Slotwinski et al. (78 – 300 K)[18]. This plot reveals that the predicted values agree with the experimental data in the 78 – 300 K range within ~5-10%. Using both sets of elastic constants as input, we calculated $E$, $G$, and the bulk modulus $B$, and plotted them as a function of temperature in Figure 4 (b,d). Details of the calculation are given in the SI. The plot in Fig. 4(b) shows an excellent agreement in $B$ between DFT and those calculated using experimental elastic constants, with an average variance of 2.51% within the entire temperature range. Figure 4c depicts the temperature dependence of $E_{<111>}$ (stiffest orientation) and $E_{<100>}$ (most compliant orientation) and shows that at 300 K, DFT predicts $E_{<111>}$ to be 21.1 GPa and $E_{<100>}$ to be 6.09 GPa; and calculation based on experimental elastic constants[18] reveal $E_{<111>}$ to be 21.2 GPa and $E_{<100>}$ to be 3.00 GPa. Figure 4(d) show the temperature dependence of $G_{<100>}$ and $G_{<111>}$ and reveals that at room temperature, the DFT predicts 8.68 GPa along <100> while calculations from experimental data yield 8.78 GPa; respective values for <111> are 2.28 GPa and 1.46 GPa. An interesting observation is that in shear, the stiffest orientation is <100>, and the most compliant is <111>, which is diametrically opposite to axial loading, with the stiffest orientation being <111> and most compliant <100>. At 440 K, DFT predicts a moderate decrease of 1.49% and 0.85 % for $G_{<100>}$ and $G_{<111>}$, respectively.

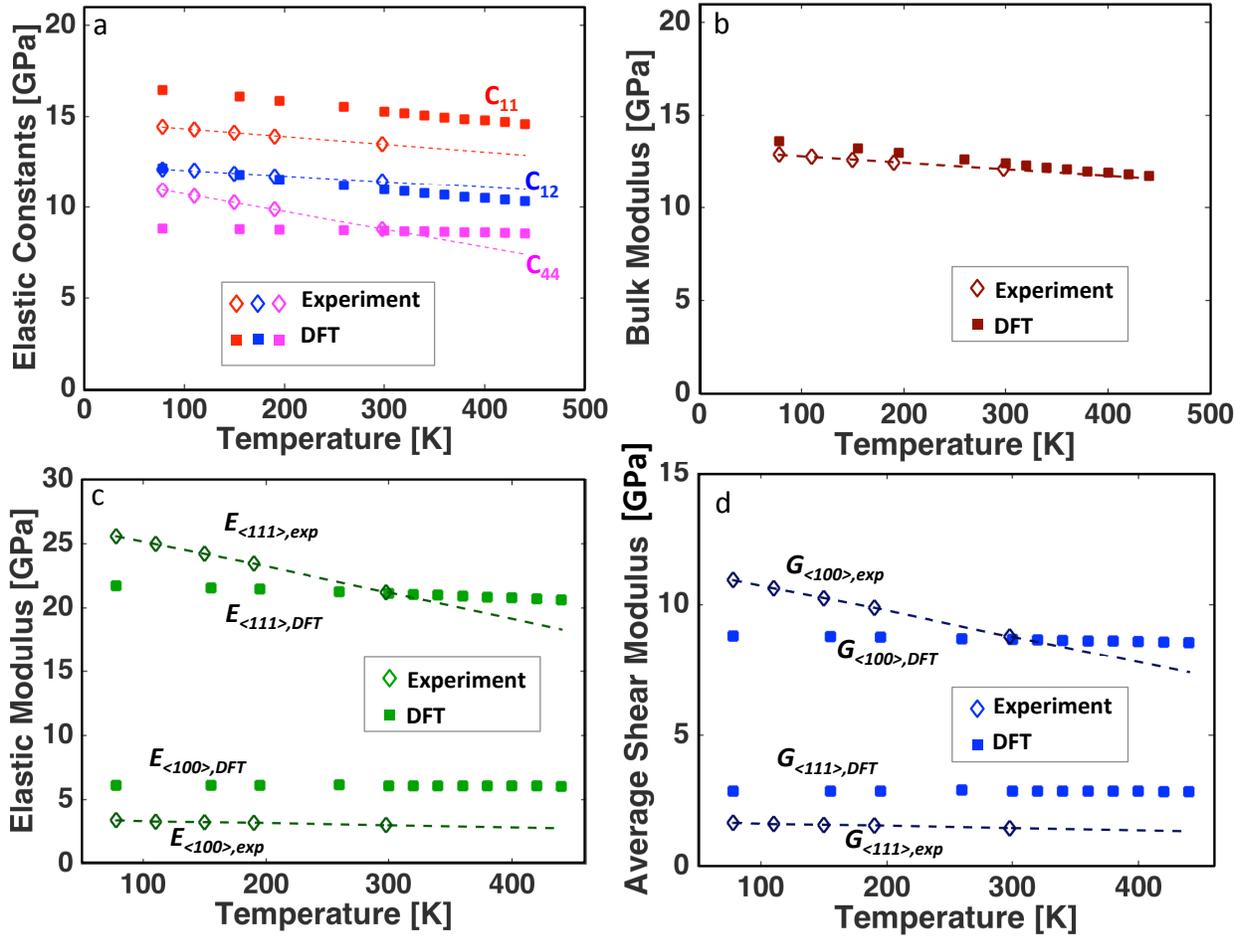

**Figure 4**. **Elastic properties of single crystalline Li at different temperatures and in different crystallographic orientations**. (a) Elastic constants $C_{11}$, $C_{12}$, $C_{44}$ as a function of temperature. Solid squares (■) represent DFT calculations (this work), diamonds (◇) represent experimental values taken from Slotwinski et al[18]. (b) Bulk modulus versus temperature. Solid squares (■) represent DFT calculations, while diamonds (◇) represent values calculated using data from Slotwinski et al.[18] (c) Maximum elastic modulus ($E_{<111>}$) and minimum elastic modulus ($E_{<100>}$) as a function of temperature as calculated by DFT and gleaned from experiments. (d) Maximum average shear modulus ($G_{<100>}$) and minimum average shear modulus ($G_{<111>}$) as a function of temperature as calculated by DFT and gleaned from experiments.

**Discussion**

The post-mortem localized deformation via crystallographic slip exhibited by Li at room temperature (Figure 1(b,e)) is similar to those reported for Nb[16,30] and V[24] nanopillars deformed at room temperature; the deformation of Mo and W nanopillars is characterized by uniform deformation with wavy slip traces[16,26,31]. A common trait shared by Nb and V is a low critical temperature ($T_c$), defined as the temperature at which screw and edge dislocations have similar mobility, and the effects of intrinsic lattice resistance become negligible[16,30]. A lack of closest-packed planes in bcc metals and the non-planarity of screw dislocation cores leads to their motion via cross-slip rather than glide-only and to their experiencing a higher effective Peierls' barrier compared with that of the edge dislocations. At higher temperatures, the thermally activated screw dislocations are able to move more easily through the potential energy landscape until $T_c$, when their mobility becomes virtually equivalent to that of edge dislocations. At that point, it becomes possible to initiate an avalanche of dislocations gliding on a single slip-plane, resulting in localized slip offsets, a common characteristic of deformation in fcc pillars[13]. $T_c$ for Nb is 350K and that for V is 380 K, which are low enough that compression at room temperature produces fcc-like deformation[32], while Mo and W have $T_c$ values of 480 K and 800 K, which leads to the waviness in the slip patterns during room-temperature deformation[16]. It is not surprising that Li would exhibit similar fcc-like deformation behavior as Nb and V, because while no report exists on the $T_c$ of Li, it is likely to be lower than its melting temperature of 453 K. In our experiments at 363 K, the post-mortem images show a transition from localized to homogeneous deformation, which is likely related to the low melting point of Li and the high homologous temperature of the experiments, $T_{test}/T_{melting}$, which is 0.65 at 298 K, and 0.8 at 363 K. At high homologous temperatures the dislocation movement is no longer confined to slip planes with the largest resolved shear stress because thermally activated processes like cross-slip and dislocation climb become possible[33]. This also explains the observed decrease in yield strength (Figure 2a,b) compared with room-temperature deformation.

The large size-effect at room temperature may be explained by one of two mechanisms: a) dislocation multiplication-driven plasticity[34,35], where the entanglement of the dislocation segments contribute to increased flow stress in a manner similar to forest-hardening in bulk crystals; or b) dislocation-nucleation governed plasticity, commonly observed in the deformation of small-scale single crystalline metals, which is accommodated by dislocation nucleation – most likely at the existing single-arm sources and/or at the surfaces – which occurs at ever higher flow stresses[32]. At 363 K, the enhancement in size effect might be caused by the increased mobility of the screw dislocations, which causes the rate of dislocation annihilation to increase and to transition overall deformation to nucleation-driven plasticity sooner. This is consistent with the size effect plot in Figure 3b, which shows that at room temperature, the power law slope of -0.68 for Li is close to those of V (-0.79) and Nb (-0.93)[24], the two bcc metals exhibiting fcc-like deformation and size-effect exponent. This plot also reveals that relative to the shear modulus, Li is the strongest of all reported bcc metals within the studied size range; for example at a pillar diameter of 1 μm ($D/b$=3300), the relative strength of Li is a factor of 1.7 higher than V and Nb, and a factor of 2.3 higher than W. At 363 K, the size-effect slope becomes -1.00, while the normalized strength decreases by a factor of ~3.5 compared to room temperature over the entire size range.

This size-effect has important implications for dendrite formation and growth that may explain why applying an external pressure higher than what is believed to be the bulk yield strength of Li at room temperature, 0.41 – 0.89 MPa[4,11], on a cell stack has not fully prevented dendrite formation[5–7]. Our findings demonstrate that during the initial stages of dendrite nucleation and growth, the yield strength of nanoscale dendrites exceeds that of the externally applied pressure, which is what allows them to grow. As the dendrites expand, their yield strength decreases to the point where it is balanced by the applied pressure, which drives them to deform plastically and promotes the in-plane growth of Li[5,7], leading to a smoother surface at the top of the deposit. The high yield strength also causes an appreciable change in the kinetics of dendrite growth through the pressure term in the modified Butler-Volmer equation[3,4]. At higher temperatures, the dramatic decrease in yield strength may explain why dendrite suppression is more easily achieved[36].

Unlike the yield strength, the elastic properties have a weak dependence on temperature. Our DFT calculations of $C_{11}$, $C_{12}$ and $C_{44}$ capture the expected asymptotic linear reduction with temperature, saturating at a finite value near the melting point, as often exhibited by other cubic crystals[37]. It is encouraging that although DFT calculations have sometimes been found to poorly predict $C_{44}$[38], our DFT prediction of $C_{44}$ for room temperature value agrees well with experimental data (1.14% difference). Figure 4c-d shows that the large difference between the strongest and weakest crystal orientation persists throughout the entire studied temperature range. Our results indicate that dendrites originating from the <100>-oriented grain in a typical polycrystalline[39] Li anode would require a separator with a shear modulus of at least 17.5 GPa to suppress them, while a value of 2.92 GPa is sufficient for the <111> orientation, according to the linear stability analysis by Monroe and Newman[3].

Our results provide insights into the mechanical properties of Li as a function of sample dimensions, temperature, and crystallographic orientation, and have significant implications for Li dendrite growth, as well as provide guidelines for their suppression. The discovered yield strength of 1 micron-sized Li of 105 MPa at room temperature represents a two-orders-of-magnitude increase over what is believed to be the bulk strength of Li, 0.41 – 0.89 MPa[4,11]., and exposes serious shortcomings of the current mechanical methods of dendrite suppression. The observed three-fold decrease in yield strength at 363 K (operating temperature of many SPEs) is substantial compared to the marginal decrease in shear modulus, which indicates that at high temperatures, dendrite suppression via inducing plastic deformation will be much more effective than finding SPEs with higher shear moduli. The high elastic anisotropy warrants the move to beyond the simple isotropic treatment of most existing theoretical efforts. More attention needs to be paid to the variation of elastic and shear moduli in the polycrystalline anode when designing SPEs with high shear modulus or when fabricating Li anodes rich in compliant orientations. Based on our experimental data and theoretical analysis, we propose that a combination of higher temperature operation together with a Li anode whose interfacial orientation has a low shear modulus, for example <111>, in contact with a solid electrolyte with high elastic modulus will prove to be an effective route to suppress dendrite formation. The rational design principles and the high-fidelity data provided will rapidly accelerate the development of a reversible Li metal anode, paving the way for higher energy density Li-ion batteries or "beyond Li-ion" chemistries such as Li-S or Li-$O_2$.

**Author Contributions**

C.X., J.R.G. and V.V. designed the research. C.X. carried out the experiments. Z.A. carried out the DFT calculations. A.A. contributed to useful discussions. V.V and J.R.G. supervised the project. All authors contributed to the interpretation of the results and to the writing of the paper.

**Acknowledgements**

The authors would like to thank Dr. Chi Ma for EBSD assistance and useful discussions. The authors gratefully acknowledge the financial support of the US Department of Energy under Contract No. DE-EE006799. V.V. acknowledges the support from the National Science Foundation CAREER award CBET-1554273.

# Enhanced strength and temperature dependence of mechanical properties of Li at small length scales and its implications for Li metal anodes

*Chen Xu[a,*], Zeeshan Ahmad[b], Asghar Aryanfar[a], Venkatasubramanian Viswanathan[b], and Julia R. Greer[a]*

[a]Division of Engineering and Applied Science, California Institute of Technology, Pasadena, CA 91125

[b]Department of Mechanical Engineering, Carnegie Mellon University, Pittsburgh, PA 15213

Address correspondence to: chenxu@caltech.edu; venkvis@cmu.edu

# Supporting Information

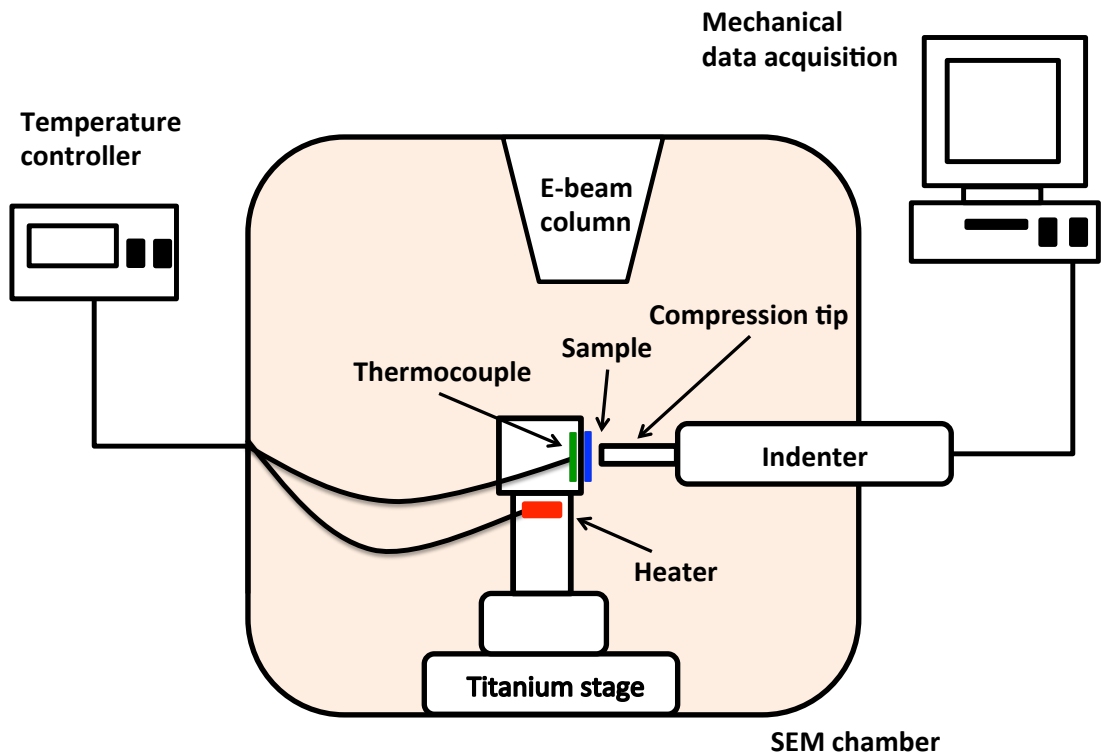

**Figure S1**. **Schematic of SEMentor**. Based on Lee et al[1]. Modifications were made to his original system to achieve above room temperature testing.

**Effect of atmospheric contamination**

To explore the effect of exposing these small-scale Li samples to atmosphere, we placed a representative 4.80 μm-diameter pillar in air for 30 seconds and then compressed it (Figure S3). The plot in Figure S3a shows that the first strain burst occurs at 50 MPa, after which the load continues to increase to ~200 MPa before catastrophic failure. The first strain burst is correlated to a pore collapsing – the Li pillar became porous and irregular after exposure to atmosphere, as can be seen on the top section of the pillar in Fig.S2 b.  The yield strength of 200 MPa is ~4 times higher than the ~50 MPa yield strength of a typical un-oxidized Li sample of similar size. The initial loading stiffness is also much higher in the oxidized sample. This is likely due to the immediate formation of LiOH, $Li_2O$ and $Li_3N$, which would have greater strength compared with Li. Post compression image shown in Figure S3b clearly reveals charging of the pillar, consistent with the formation of insulating oxides through a significant portion of the pillar.

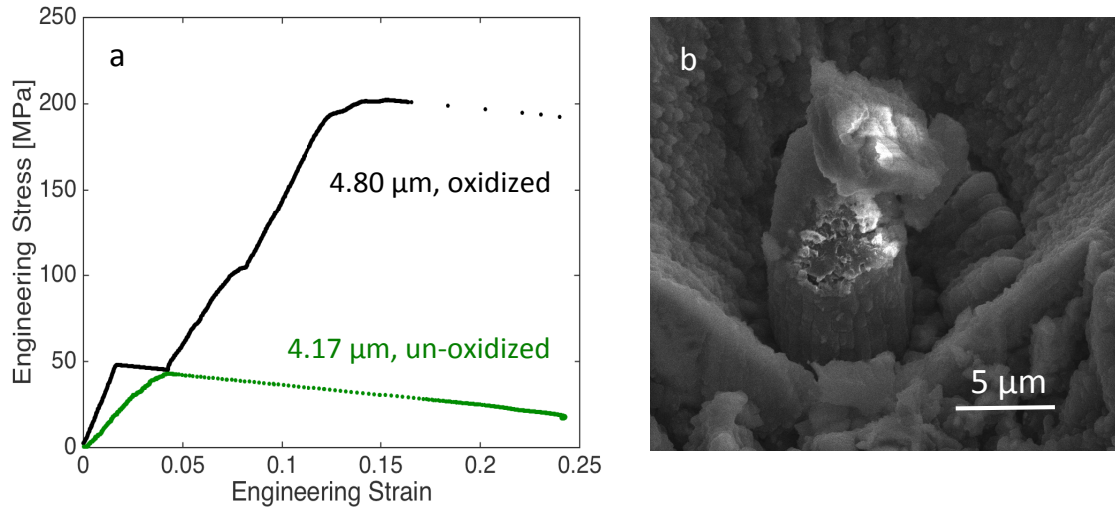

**Figure S2. Compression results of oxidized an un-oxidized pillar**. (a) Engineering stress vs. engineering strain data of the compression of a 4.8 μm oxidized pillar, comparing with the data of a 4.17 μm un-oxidized pillar. The 1st strain burst during loading is due to cracks formed via oxidation. (b) Post-compression image of the oxidized pillar. White parts due to the charging effect of non-conductive material.

**Identifying the crystal orientation of pillars**

We show the grain structure of melted and then subsequently cooled down Li metal in Figure S4a. EBSD was conducted on a blade-cut surface showing a mirror finish. Chemical polishing has been unsuccessful in further smoothing the surface. Nash *et al.*[4] found that methanol was the best etchant for Li metal, and that repeated actions of etching in methanol for a few seconds, then rinsing in xylene would allow for visual observation of grain boundaries. However this process must be done in the open air. If the same procedure was followed in an Ar atmosphere, a white crust forms on the surface of Li as soon as it is removed from methanol. This is exactly what we have found in our studies. Ethanol and isopropanol was also used to no avail. Preparation of Li metal outside the glovebox, though it may allow for visual observation of grain boundaries, introduces an oxide layer that makes it impossible to characterize the underlying Li using EBSD. Therefore EBSD was conducted on an as-cut lithium metal surface, where the roughness prohibited indexing of certain regions. Regardless, the indexed points are sufficient in indicating the average grain size of our lithium sample, around 250±86 μm, as shown in Figure S4 (a,b). The fact that our pillars, with diameters less than 10 μm, are much smaller than the average grain size leads us to conclude that our pillars are single crystalline. This assumption is often used in the nanomechanical community when studying polycrystalline material[5]. An instance of pillars lying within a single grain this is shown in Figure S4c. In some instances however it is difficult to identify which particular grain the pillars belong to, as shown in Figure S4d. In these cases, in order to obtain the crystal orientation of pillars to calculate the Schmidt factor, we utilized the relation between the elastic modulus of a particular grain and its miller indices as given below[6]

$$\frac{1}{E_{hkl}} = s_{11} - 2(s_{11} - s_{12} - \frac{1}{2}s_{44}) \cdot \frac{h^2k^2 + h^2l^2 + k^2l^2}{(h^2 + k^2 + l^2)^2} \tag{1}$$

where $h$, $k$ and $l$ are the miller indices of the respective grain, $S_{11}$, $S_{12}$ $S_{44}$ are the elements of the compliance matrix, and $E_{hkl}$ is the orientation-dependent elastic modulus. $E_{hkl}$ is obtained from the uniaxial compliance measured from unloading. Elements of the compliance matrix are calculated from the elastic constants[4,7,8]. The symmetry of Equation (1) prevents us from uniquely identifying the exact sequence and sign of the $h$, $k$ and $l$ values, however the maximum Schmidt factor is the same for every permutation.

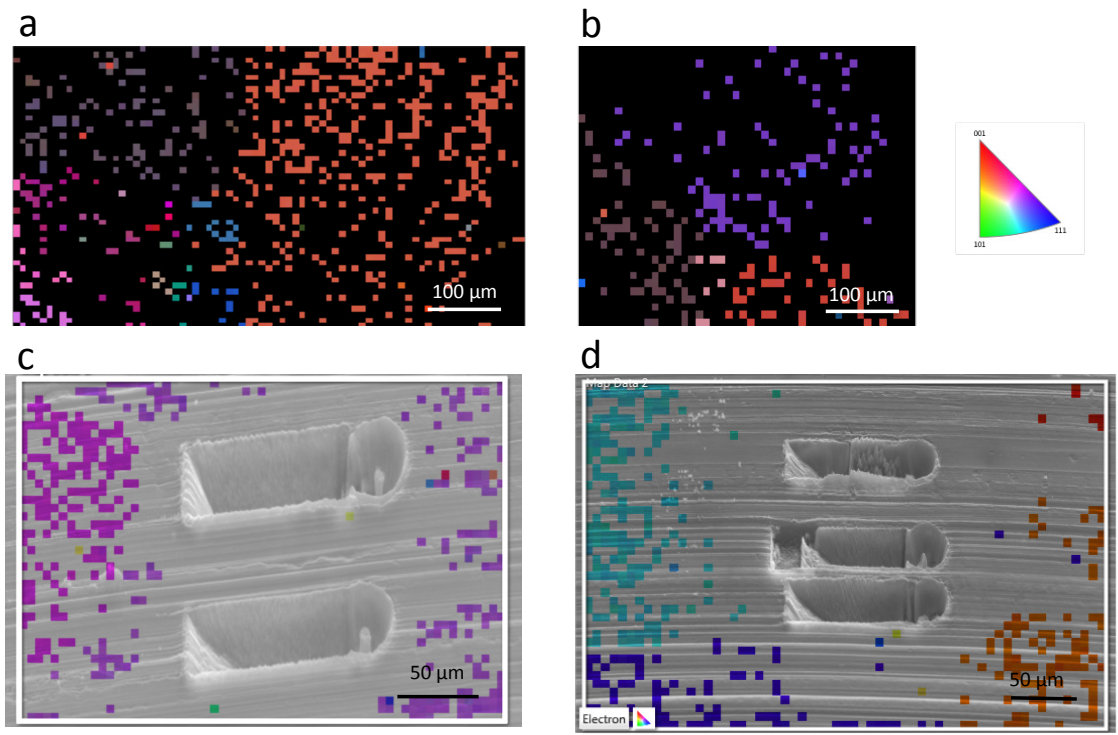

**Figure S3. Crystal grain map of Li substrate with pillars post compression.** (**a,b**) Orientation Imaging Microscopy map generated by EBSD showing the annealed and as-cut surface of the Li sample, with grain size 250±86 μm (**c,d**) SEM image of pillars post-compression overlaid with orientation mapping.

**Effect of Ga⁺ implantation**

Indentations are performed on FIB'd and as-cut Li surfaces. Ion beam with a 30nA current was used to polish the surface of as-cut Li at a grazing angle. We performed nanoindentations to a depth of up to 3 μm to match the deforming indentation volume to that of the pillars. The load vs. displacement data is for the FIB-polished and as-cleaved Li surfaces is shown in Figure S2 and reveals that these surface treatments lead to statistically indistinguishable mechanical properties. These results are consistent with multiple previous reports, which demonstrated that FIB-irradiation did not significantly affect the strength and deformation of metallic micro- and nano-pillars[2,3].

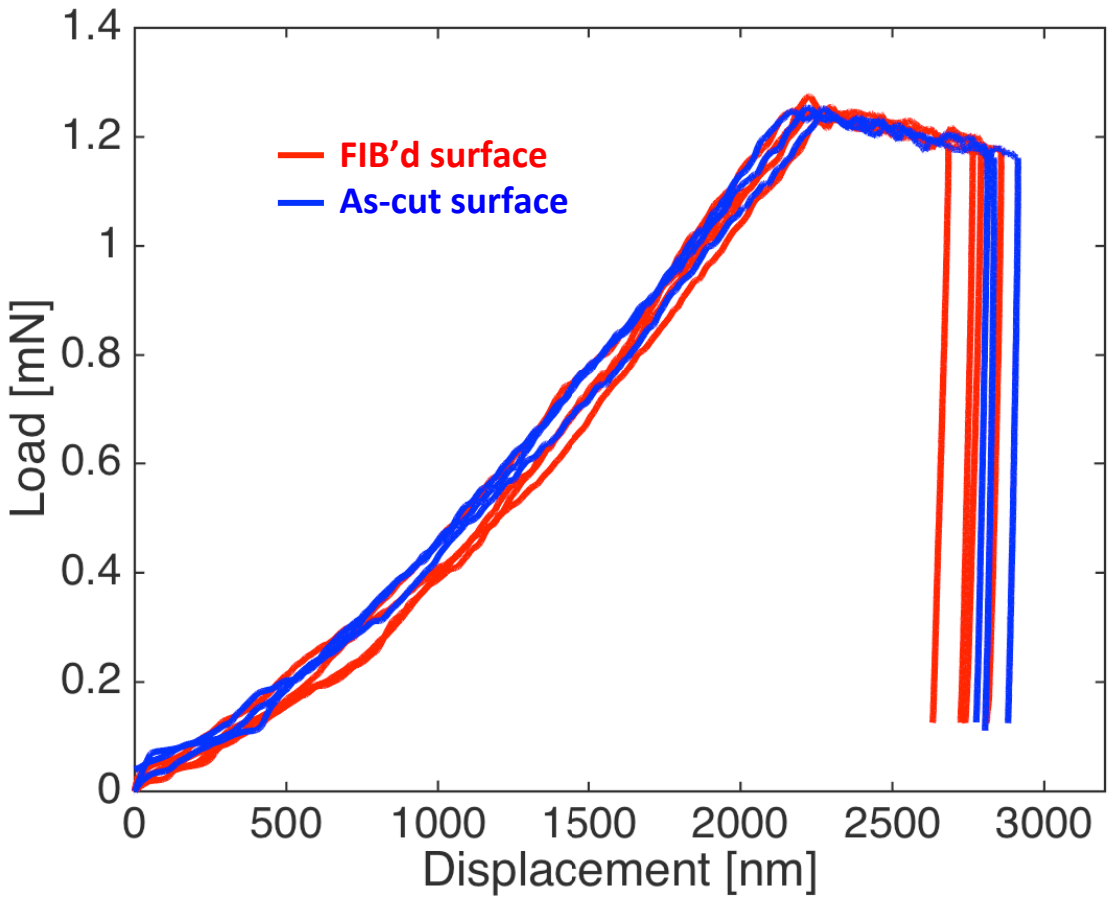

**Figure S4.** Load controlled indentation of Li metal with a cleaved surface (red) and FIB polished surface (blue). The conditions for the FIB are 30keV acceleration voltage and 30nA current.

**Calculating the average shear modulus of a single crystal**

Turley and Sines[9] gave the general expression for the shear modulus of a single crystal below

$$G = \frac{1}{s_{44} + 4(s_{11} - s_{12} - \frac{1}{2}s_{44})\Omega} \quad (1)$$

where $s_{11}$, $s_{12}$ and $s_{44}$ are elements of the compliance matrix, and $\Omega$ is the angle dependence as defined below

$$\Omega = a + b\sin 2\theta + c\cos 2\theta \quad (2)$$

Where $a$, $b$ and $c$ are linear combinations of the directional cosines of (*hkl*). It is clear that when Eq. (2) is averaged over $2\pi$ with respect to $\Theta$, the sinusoidal terms disappear. The expression for $a$ is given as

$$a = A^2 B^2 + C^2 D^2 \quad (3)$$

where

$$\begin{aligned} A &= \cos\alpha\cos\beta \\ B &= \sin\alpha\cos\beta \\ C &= \sin\beta \\ D &= \cos\beta \end{aligned} \quad (4)$$

and

$$\sin\alpha = \frac{k}{(h^2+k^2)^{1/2}}; \quad \sin\beta = \frac{l}{(h^2+k^2+l^2)^{1/2}}$$

$$\cos\alpha = \frac{h}{(h^2+k^2)^{1/2}}; \quad \cos\beta = \left(\frac{h^2+k^2}{h^2+k^2+l^2}\right)^{1/2}$$

(5)

A schematic is provided in Figure S5.

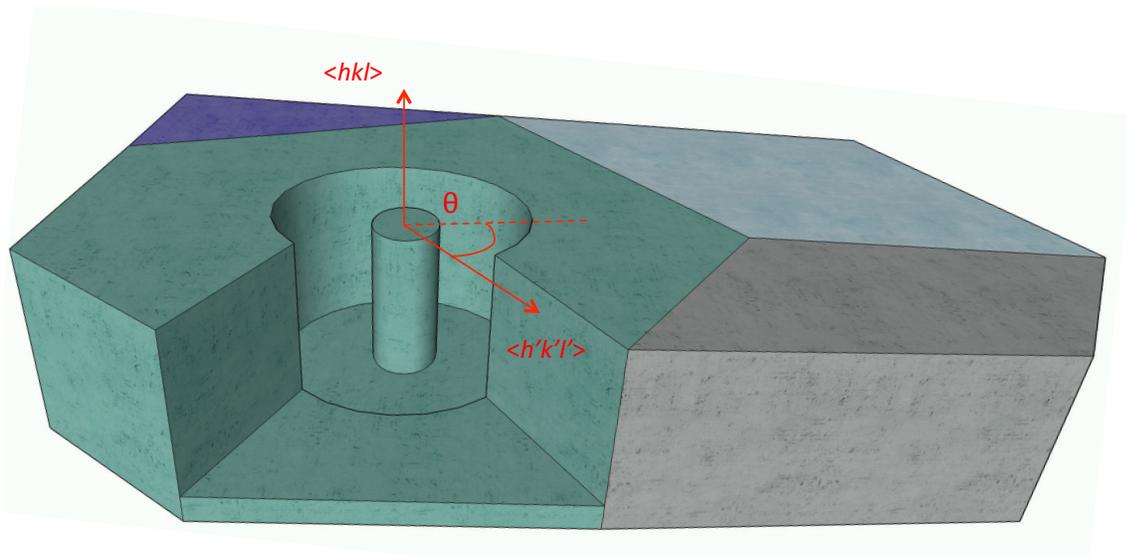

**Figure S5**. A single crystalline pillar situated within a crystal grain with normal in the direction of *<hkl>*. The different colored regions indicate different crystal grains imbedded in a polycrystalline Li metal foil. *<h'k'l'>* is a transverse direction in the (*hkl*) plane. Θ is the angle between *<h'k'l'>* and the Meridional tangent (indicated by the dashed line). For the definition please refer to Turley et al.[9]

**Density Functional Theory (DFT) Calculations**

The DFT were performed using a real space projector augmented wave (PAW)[10] method as implemented in GPAW[11] within the generalized gradient approximation. A real space grid spacing of 0.18 Å and Monkhorst Pack[12] scheme for k-points consisting of 3456 k-points per reciprocal atom or higher was used. All calculations were converged to the energy of < 0.1 meV and a force of < 0.01 eV/Å. The phonon vibrational frequencies and the density of states were calculated using the small displacement method[13] as implemented in Atomic Simulation Environment (ASE)[14]. A 32-atom supercell of lithium was used for all phonon calculations. The lattice constant at each temperature was calculated by free energy minimization and the vibrational contribution to free energy was calculated at each of the temperatures using the method described below.

We express the free energy $F$ in terms of the strain state at a constant volume as[15,16]:

$$F = F_0 + \frac{V}{2}\sum_{i=1}^{6}\sum_{j=1}^{6} C_{ij} e_i e_j + O(e_i^3) \quad (6)$$

where $F_0$ denotes the free energy at zero strain. We used volume-conserving orthorhombic strain and monoclinic strains for cubic crystals[15]. The volume-conserving orthorhombic strain is given by:

$$\begin{aligned} e_1 &= -e_2 = x, \\ e_3 &= \frac{x^2}{1-x^2} \\ e_4 &= e_5 = e_6 = 0. \end{aligned} \quad (7)$$

In this case, the free energy change is an even function of the strain, and is given by:

$$F = F_0 + V(C_{11} - C_{12})x^2 + O(x^4). \tag{8}$$

This allows the determination of the elastic constant $C_{11} - C_{12}$. In a similar fashion, volume-conserving monoclinic strain is given by:

$$e_6 = x$$
$$e_3 = \frac{x^2}{4 - x^2} \tag{9}$$
$$e_1 = e_2 = e_4 = e_5 = 0.$$

In this case, the resulting free energy change is again an even function of the strain given by:

$$F = F_0 + \frac{V}{2}C_{44}x^2 + O(x^4). \tag{10}$$

This allows us to calculate the elastic constant $C_{44}$.

To isolate the individual elastic constants, $C_{11}$ and $C_{12}$, we used the relationship $B = (C_{11} + 2C_{12})/3$, where B is the bulk modulus, calculated by fitting Birch-Murnaghan equation of state to the free energy variation with volume obtained from the density functional theory calculations[17].

We account for the free energy contribution due to lattice vibrations using the quasi-harmonic approximation, where the total free energy for a cubic crystal can be written as[18,19]:

$$F(V,T) = E(V) - k_B T \ln Z(V,T) \tag{11}$$

where $E$ is the energy of the frozen lattice at volume $V$, $k_B$ is the Boltzmann constant and $Z$ is the partition function associated with the vibrations at temperature $T$.[18] We consider only the phonon vibrational contributions, which can be expressed in terms of the phonon vibrational frequencies $\omega$ and density of states $g(\omega, V)$, given by:

$$Z(V,T) = \int_0^\infty g(\omega, V) \ln \frac{1}{2 \sinh\left(\frac{\hbar\omega}{k_B T}\right)} d\omega \qquad (12)$$